\documentclass[prd,showpacs,superscriptaddress,preprintnumbers,twocolumn,footinbib,amsmath,showkeys]{revtex4}
 \usepackage[dvips,final]{graphicx}
  \usepackage{amssymb,amsmath,epsfig,bm,pifont}
   \graphicspath{{./figs/}}
\usepackage{color}

\definecolor{GrayW}{rgb}{0.50196,0.50196,0.50196}
 \newcommand{\GrayW}[1]{{\color{GrayW} #1}}
\usepackage{relsize}
\newcommand{\babar}{{\mbox{\slshape B\kern-0.1em{\smaller A}\kern-0.1em
            B\kern-0.1em{\smaller A\kern-0.2em R}}}
           }
\usepackage[utf8]{inputenc}
\def\MSbar{\relax\ifmmode\overline
            {\rm MS}\else{$\overline{\rm MS}${ }}\fi}
\newcommand{\Ds}{\displaystyle}                           

\begin{document}
\thispagestyle{empty}
 \date{\today}
  \preprint{\hbox{RUB-TPII-02/2021}}

\title{Cross-link relations between $\pi$ and $\rho$-meson channels and the QCD vacuum\\}

\author{S.~V.~Mikhailov}
\email{mikhs@theor.jinr.ru}
\affiliation{Bogoliubov Laboratory of Theoretical Physics, JINR,
             141980 Dubna, Russia}

\author{N.~G.~Stefanis}
\email{stefanis@tp2.ruhr-uni-bochum.de}
\affiliation{Ruhr-Universit\"{a}t Bochum,
             Fakult\"{a}t f\"{u}r Physik und Astronomie,
             Institut f\"{u}r Theoretische Physik II,
             D-44780 Bochum, Germany}

\date{\today}
\begin{abstract}
We discuss cross-link relations between the $\pi$ and $\rho$-meson
channels emerging from two different descriptions of the QCD vacuum:
Instanton physics and QCD sum rules with nonlocal condensates (NLC).
We derive in both schemes an intriguing linear relation between the
$\pi$ and the $\rho^\|$-meson distribution amplitudes in terms of their
conformal coefficients and work out the specific impact of the scalar
NLC in these two channels.
Using a simple model with Gaussian decay of the scalar NLC, we are able
to relate it to the moments of the pion non-singlet parton distribution
function measurable in experiment---a highly nontrivial result.
The implications for the pion and the $\rho^\|$-meson DAs entailed by
the obtained cross-link relations are outlined in terms of two generic
scenarios.
\end{abstract}
\pacs{12.38.Aw,12.38.Lg,14.40.Be}
%
\keywords{pion and rho-meson distribution amplitudes,
          pion quark distribution function,
          nonlocal condensates,
          instanton vacuum
          }
\maketitle

\section{Introduction}
\label{sec:intro}
Recently, Polyakov and Son \cite{Polyakov:2020cnc} have employed the
instanton approach in combination with dispersion relations for the
two-pion distribution amplitude (DA) \cite{Polyakov:1998ze} to
obtain model-independent relations for the ratio
$a_{2}^{(\rho)}/a_{2}^{(\pi)}$.
Within this approach, the Gegenbauer moments of the $\rho$-meson DA can
be expressed in terms of the two-pion distribution amplitudes (2$\pi$DAs).
Using the conformal expansion
\begin{equation}
  \varphi_{\pi(\rho)}^\text{(tw-2)}(x,\mu^2)
=
  \psi_{0}(x) + \sum_{n=2,4,\ldots}^{\infty} a_{n}(\mu^2) \psi_{n}(x)
\label{eq:DA-conf-exp}
\end{equation}
with the Gegenbauer basis
\begin{equation}
  \psi_n(x)
=
  6x\bar{x}\,C_{n}^{(3/2)}(x-\bar{x}) \, ,
\label{eq:harmonics}
\end{equation}
where
$\varphi_{\pi}^\text{asy}=\psi_{0}(x)=6x(1-x)\equiv 6x\bar{x}$
is the asymptotic pion DA,
it was found that $a_{2}^{(\rho)}$ and $a_{2}^{(\pi)}$ are
linearly related.
It was argued that this is deeply rooted in chiral dynamics and the
general properties of quantum field theory (QFT), such as unitarity,
crossing, and dispersion relations.
Going one step further, they used the soft pion theorem and the
crossing symmetry \cite{Polyakov:1998ze} to relate the second
Gegenbauer coefficient $a_{2}^{(\rho)}$ of the $\rho$-meson DA to
the third Mellin moment of the pion valence-quark parton distribution
function (PDF) measured in DIS.
Referring for the full derivation to \cite{Polyakov:2020cnc}, one obtains
the following relation
\begin{eqnarray}
  a_{2}^{(\rho)}
& = &
  B_{21}(0) \exp{\left(c_{1}^{(21)}m_{\rho}^{2}\right)} \nonumber \\
& = &
  \left(
        a_{2} ^{(\pi)} - \frac{7}{6}M_{3}^{(\pi)}
  \right)
  \exp{\left(c_{1}^{(21)}m_{\rho}^{2}\right)} \, .
\label{eq:second-moments}
\end{eqnarray}

The main ingredient in the last equation is the measurable quantity
\begin{equation}
  M_{3}^{(\pi)}
=
  \int_{0}^{1} dx x^2 \left[q_\pi(x) - \bar{q}_\pi(x)\right] \, ,
\label{eq:2moment-pion-QDF}
\end{equation}
while $c_{1}^{(21)}$ is a low-energy subtraction constant in the
dispersion relation for the generalized Gegenbauer moments
$B_{nl}(W^2)$ used for the 2$\pi$DAs in \cite{Polyakov:1998ze}
(see there for details).
Fixing this constant by virtue of the instanton model of the QCD
vacuum, one can determine the ratio $a_{2} ^{(\rho)}/a_{2} ^{(\pi)}$.
To this end, we combine the value
\begin{equation}
  a_{2}^{(\pi)}(\mu=2~\text{GeV})
= 0.078\pm 0.028 \, ,
\label{eq:a2_pi-lattice}
\end{equation}
determined on the lattice with a next-to-leading order (NLO) matching to the \MSbar scheme
(adding the errors in quadrature) \cite{Bali:2019dqc},
with the value in the first line below
\begin{eqnarray}
  M_{3}^{(\pi)}(\mu=2~\text{GeV})
= \bigg\{
		\begin{array}{l}
			 0.114 \pm 0.020~~\text{\cite{Novikov:2020snp}}
\label{eq:M3} \\
			 0.110(7)(12)~~~\text{\cite{Alexandrou:2021mmi}}\, ,
		\end{array}
\end{eqnarray}
obtained at NLO in the phenomenological analysis in \cite{Novikov:2020snp}
within the xFitter framework.
The second line shows the result of a recent lattice calculation
\cite{Alexandrou:2021mmi}.
One observes that the lattice estimate is within the error margin of the
first line.
This way one gets \cite{Polyakov:2020cnc}
\begin{equation}
  \frac{a_{2}^{(\rho)}}{a_{2}^{(\pi)}}(\mu=2~\text{GeV})
=
  -(1.15\pm 0.86)(1.0\pm 0.1) \, .
\label{eq:ratio}
\end{equation}

The above results provide the motivation for the present investigation
with the attempt to derive similar linear relations between the $\pi$ and
$\rho$-meson channels within the QCD sum-rules approach with nonlocal
condensates (NLC-SR for short) used in \cite{Bakulev:2001pa}.
The key ingredients in NLC-SR are the nonlocal condensates
\cite{Mikhailov:1986be,Mikhailov:1988nz,Mikhailiov:1989mk,
Bakulev:1991ps,Mikhailov:1991pt}.
In particular the scalar NLC $\Phi_{S}(x,M^2)$,
where $M^2$ is the Borel parameter, and the scale
\begin{equation}
  \lambda_{q}^{2}
=
  \frac{\langle\bar{q}D^{2}q\rangle}
  {\langle\bar{q}q\rangle}\Big|_{\mu_0^2\thicksim 1\text{GeV}^2}
=
  0.40 \pm0.05~\mbox{GeV}^2 \, ,
\label{eq:quark-virtuality}
\end{equation}
which defines the virtuality of vacuum quarks,
are crucial for the determination of the shape of the pion DA
employing NLC-SR \cite{Bakulev:2001pa,Bakulev:2002hk}.

This approach was originally used to derive a set of pion DAs
(termed Bakulev-Mikhailov-Stefanis or BMS) for
$\lambda_{q}^{2}=0.40$~GeV$^2$ in agreement with the
CLEO experimental data \cite{Gronberg:1997fj} on the pion-photon
transition form factor (TFF).
It also provides good agreement with more recent
data from Belle \cite{Uehara:2012ag} and
\textit{BABAR}($Q^2\leqslant 9$~GeV$^2$) \cite{Aubert:2009mc},
see \cite{Stefanis:2020rnd}.
The coefficients
$a_{2}^{(\pi)}$ and $a_{4}^{(\pi)}$
encapsulate the main theoretical characteristics of
the nonperturbative QCD vacuum.
Later, this scheme was employed in \cite{Stefanis:2014nla}
to construct another variant of the pion DA with a platykurtic profile.
This DA intrinsically combines
two key features: endpoint suppression of the pion DA due to
the vacuum quark virtuality $\lambda_{q}^{2}=0.45$~GeV$^2$
and unimodality of the DA at $x=1/2$, as found in the context of
Dyson-Schwinger equations to account for the dynamically generated mass
of the confined quark propagator, see \cite{Roberts:2021nhw} for a review.
At the scale $\mu^2=4$~GeV$^2$ one has the coefficients
$(a_{2}^{(\pi)}, a_{4}^{(\pi)})^{\mu^2}_\text{pk}=(0.057,-0.013)$.
This scheme \cite{Stefanis:2015qha} also provided a platykurtic DA
for the longitudinally polarized $\rho$-meson:
$(a_{2}^{(\rho)}, a_{4}^{(\rho)})_\text{pk}^{\mu^2}=(0.017,-0.021)$,
while previous attempts to construct the $\rho$-meson DA within this
NLC-SR scheme were reported in \cite{Bakulev:1998pf,Pimikov:2013usa}.

The paper is structured as follows.
In Sec.\ \ref{sec:rho-pi-SR} we compare the predictions for the
$\pi$-$\rho$-meson coefficients following from
Eq.\ (\ref{eq:second-moments}) in \cite{Polyakov:2020cnc}.
We then derive an analogous expression to (\ref{eq:second-moments}) using
the NLC-SR scheme and discuss the dynamical difference of the impact of
the scalar NLC in the axial and the vector channel.
We also work out a linear relation between functionals depending on
$\varphi_{\rho}^{\|}$ and $\varphi_\pi$ and extend the result to
higher orders of the conformal expansion.
In Sec.\ \ref{sec:NLC-M3}, we establish the linear
relation between the $\pi$ and $\rho$-meson DAs in terms of the specific
characteristics of the NLC-SR.
The ultimate goal of this section is to relate the scalar NLC to the
third/fifth Mellin moment of the pion PDF.
The implications of the negative sign of $a_2^{(\rho)}$ for the pion
and $\rho^\|$-meson DAs are addressed in
Sec.\ \ref{sec:implications-pion-DA}.
Our conclusions are presented in Sec.\ \ref{sec:concl}.
The expressions for the scalar NLC are provided in Appendix \ref{app:A},
while the correspondence between the instanton approach and the NLC-SR
scheme is shown in Appendix \ref{app:B}.

\section{How to relate the $\rho^\|$ and $\pi$ distribution amplitudes in NLC-SR}
\label{sec:rho-pi-SR}
\begin{figure}[t]
\includegraphics[width=0.45\textwidth]{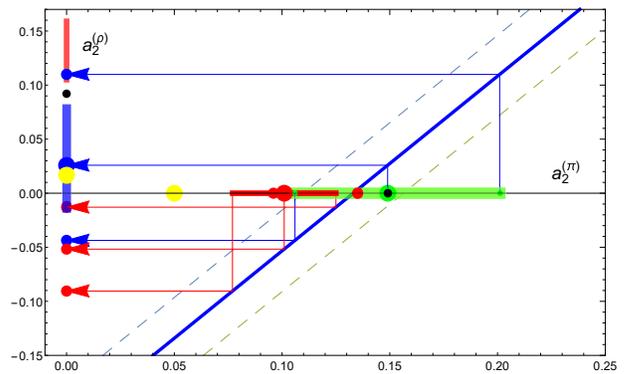}
\caption{Comparison of domains of pairs
$\left(a_{2}^{(\pi)}, a_{2}^{(\rho)}\right)$ obtained in different
approaches at $\mu^2=4$ GeV$^2$ obeying the linear relation
given by Eq.\ (\ref{eq:second-moments}).
The blue diagonal line and the dashed lines around it originate from
this relation and from the uncertainties of $M_{3}^{(\pi)}$
in (\ref{eq:2moment-pion-QDF}).
Lattice constraints for $a_2^{(\pi)}$ at NNLO
provide the range $[0.077,0.125]$---shown as a red segment
with a central big red point along the horizontal axis.
The N$^3$LO regime corresponds to the interval
$[0.096, 0.135]$ \cite{Bali:2019dqc}
limited by the red endpoints.
The range $a_{2}^{(\rho)}=0.132\pm 0.027$ from \cite{Braun:2016wnx}
is shown as a short red segment at the upper end of the vertical axis.
The NLC-SR fiducial regions are shown for the pion by the green segment
\cite{Bakulev:2001pa} and for the $\rho^\|$-meson by the blue segment
along the vertical axis \cite{Bakulev:1998pf}.
The projections of the endpoints of the green segment and its
center (green dot) are denoted by blue points on the $a_{2}^{(\rho)}$-axis.
The yellow points mark the locations of the platykurtic pion and $\rho$
DAs \cite{Stefanis:2014nla,Stefanis:2015qha}, respectively.
The black points $a_2^{(\pi)}=0.149$
and $a_2^{(\rho)}=0.092$ denote the coefficients of the DSE DAs for
$\pi$ \cite{Chang:2013pq}
and $\rho^\|$ \cite{Gao:2014bca}.
\label{fig:a2-comparison}}
\end{figure}

It is instructive to compare the predictions for pairs
$\left(a_{2}^{(\pi)}, a_{2}^{(\rho)}\right)^{\mu^2}$,
obtained within various approaches, with the cross-link relations
following from Eq.\ (\ref{eq:second-moments}).
This is done in Fig.\ \ref{fig:a2-comparison} in terms of the blue
diagonal line which represents the linear relation between such pairs.

The following results are displayed. \\
i)   $a_2^{(\pi)}$ values of the set of BMS DAs
     \cite{Bakulev:2001pa}: green segment. \\
ii)  $a_{2}^{(\rho)}$ range from NLC-SR in \cite{Bakulev:1998pf}:
     blue segment bounded by blue points in good agreement with the
     projections of the endpoints of the green $a_2^{(\pi)}$ segment. \\
iii) $a_2^{(\pi)}$ estimates from lattice QCD with NNLO and N$^3$LO
     matching to the \MSbar scheme \cite{Bali:2019dqc}: red segment
     and interval between the smaller red endpoints, respectively. \\
iv)  $a_{2}^{(\rho)}$ from the lattice calculation
     in \cite{Braun:2016wnx}: red segment at the upper end of the vertical axis. \\
v)   NLC-SR platykurtic $\pi$ and $\rho^\|$ DAs
     \cite{Stefanis:2014nla,Stefanis:2015qha}: yellow dots
     on the horizontal and vertical axis, respectively. \\
vi)  The black dots within the green $a_2^{(\pi)}$ and blue $a_2^{(\rho)}$
     intervals show the locations of the pion \cite{Chang:2013pq} and the
     $\rho^\|$-meson \cite{Gao:2014bca} DAs, respectively, obtained from a
     Dyson-Schwinger equations (DSE) based approach---\cite{Chang:2013pq}
     (pion) and \cite{Gao:2014bca} ($\rho$-meson).
     One notices that the
     $a_2^{(\pi)}\rightarrow a_2^{(\rho)}$ projection
     and the original $a_2^{(\rho)}$ value do not coincide.

A key observation from this figure is the existence of a boundary for the
pion coefficient $a_2^{(\pi)}=A_2^{(\pi)}$ below which the corresponding
$\rho^\|$ coefficient $a_2^{(\rho)}$ becomes negative.
Using the estimates for the parameters $M_3, c_1^{(21)}$ given above,
cf.\ (\ref{eq:M3}), we find
\begin{equation}
  A_2^{(\pi)}
\approx
  0.1
\label{eq:boundary}
\end{equation}
in agreement with Eq.\ (\ref{eq:ratio}).

Let us also emphasize the good correspondence in i) and ii) between
the segments of $a_2^{(\pi)}$ and $a_2^{(\rho)}$ derived from NLC-SR.
This circumstance makes it tempting to derive a relation similar to
Eq.\ (\ref{eq:second-moments}) within the NLC-SR approach for the pion
\cite{Bakulev:1998pf,Bakulev:2001pa} and the $\rho^\|$-meson
\cite{Bakulev:1998pf,Pimikov:2013usa,Stefanis:2015qha}.
We start with the generalized sum rules for the $\pi$ and $\rho$-meson
channels.
Taking recourse to the sum rules for the pion in the axial channel
and for the $\rho$-meson in the vector channel \cite{Bakulev:1998pf}
(equations (7) and (8) there), we write
\begin{eqnarray}
 \!\!\varphi_{\rho}^{\|}(x)
\!\!&&\!\! =
  \left[
        \varphi_\pi(x)\! +\! \Delta_{A_{1}\rho'}(x, M^2)
        - \frac{2}{f_{\pi}^2} \Phi_S(x,M^2)\! \right] \nonumber \\
&& \times
        e^{C(M^2) m_{\rho}^2}
\label{eq:rho-pi-DAs}
\end{eqnarray}
with $f_\pi\approx 0.132$GeV, $f_\rho\approx 0.21$GeV,
\begin{equation}
  C(M^2)
=
  \frac{1}{M^2} + \frac{1}{m_{\rho}^2}\ln \left(f_{\pi}^2/f_{\rho}^2\right)\, .
\label{eq:C}
\end{equation}
The term $\Delta_{A_{1}\rho'}$ is determined by the difference
of the contributions of the higher
resonances in the phenomenological parts of the QCD SR
for the axial and vector channels:
\begin{eqnarray}
&&  \Delta_{A_{1}\rho'}(x, M^2)
=  \, \label{eq:Delta-A1}\\
&&\!\!\! \left(\frac{f_{A_1}}{f_\pi}
  \right)^2
            e^{-m_{A_1}^2/M^2} \varphi_{A_1}(x)
- \left(\frac{f_{\rho'}}{f_\pi}
  \right)^2
            e^{-m_{\rho'}^2/M^2} \varphi_{\rho'}(x).
\nonumber
\end{eqnarray}
Here $M^2$ is the Borel parameter within a stability window to be
determined later in accordance with the standard QCD SR practice
\cite{Shifman:1978bx} in such a way as to best reproduce the DA moments.
Strictly speaking, Eq.\ (\ref{eq:rho-pi-DAs}) should be considered
\textit{in the weak sense}, i.e., for smooth convolutions on \textit{both}
sides within the stability domain of $M^2$,
see, for instance, \cite{Bakulev:2001pa,Bakulev:1998pf}.

\begin{figure}[h]
\includegraphics[width=0.45\textwidth]{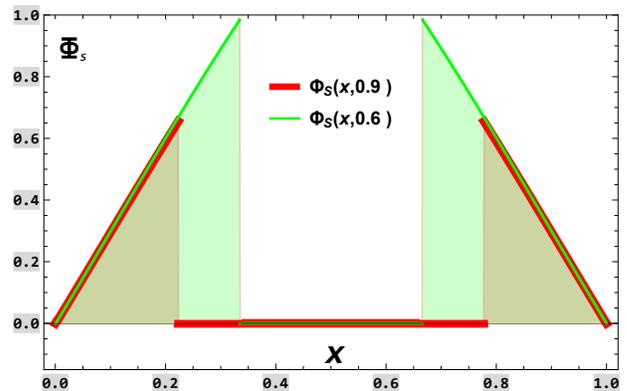}
\caption{
Scalar condensate $\Phi_S(x,M^2)$ as a function of $x\in[0,1]$
using the value $\lambda_{q}^{2}=0.4$~GeV$^2$, see \cite{Bakulev:2001pa}.
The lighter shaded areas correspond to $M^2=0.6$~GeV$^2$ and the
darker shaded areas to $M^2=0.9$~GeV$^2$.
An analogous notation is used for the thinner (green)
and thicker (red) lines.
\label{fig:scalar-cond}}
\end{figure}

The expression for $\Phi_S(x,M^2)$ together with some
explanations is given in Appendix \ref{app:A}
(see also \cite{Bakulev:2001pa,Mikhailov:2021znq}),
while the decay constants of the next resonances have the values
\begin{equation}
  f_{A_1} \approx 0.21\text{GeV}, ~~~ f_{\rho'} \approx 0.175\text{GeV}
\label{eq:decay-const}
\end{equation}
and have been determined in \cite{Bakulev:2001pa} and
\cite{Bakulev:1998pf}, respectively.
The dependence of the scalar NLC $\Phi_S(x,M^2)$
on the Borel scale $M^2$ and the quark virtuality
$\lambda_{q}^{2}=0.4$~GeV$^2$, is given graphically in
Fig.\ \ref{fig:scalar-cond} making use of the simplest Gaussian
model for the QCD vacuum from \cite{Mikhailov:1988nz,Bakulev:2002hk}.

Let us now consider the origin of the difference between
$\varphi_{\rho}^{\|}$ and $\varphi_\pi$ ensuing from the particular
structure of Eq.\ (\ref{eq:rho-pi-DAs}).
The term $\Delta_{A_{1}\rho'}$ represents a simple phenomenological
contribution that takes into account the difference of the higher
resonances in the vector and axial channels as expressed by their decay
constants, cf.\ (\ref{eq:Delta-A1}).
In contrast, the appearance of the term
$\Ds \frac{2}{f_{\pi}^2} \Phi_S(x,M^2)$
represents an evident dynamical manifestation of the distinct character
of interaction of the scalar vacuum condensate in these two channels.
Moreover, it is usually the dominant contribution.

What is the physical reason for the dynamical difference,
encoded in $-2\,\Phi_S$, between
$\varphi_{\rho}^{\|}$ and $\varphi_\pi$\,
in Eq.\ (\ref{eq:rho-pi-DAs})?

The pion DA $\varphi_\pi$ is extracted from the axial-axial correlator
because it originates as the pion projection of the \textit{axial}
(nonlocal) current, while the $\varphi_{\rho}^{\|}$ DA is extracted from
the vector-vector correlator because it is the $\rho$-meson projection
of the (nonlocal) \textit{vector} current.

The scalar NLC is the only vacuum condensate,
among others included in the used approximation,
whose contribution is affected by the gamma-matrix structure of the
coefficient function.
All other condensate contributions to the theoretical part of the QCD SR
are the same for both channels, i.e., axial-axial and vector-vector, of
these correlators and are accumulated in $\varphi_\pi$.
As a result, the repeated commutations of $\gamma_5$ with an axial vertex,
change the sign of the coefficient function of the scalar condensate
relative to a chain of commutations with vector vertices.
We will discuss the implications of these effects for the $\rho$ and $\pi$
DAs in Sec.\ \ref{sec:implications-pion-DA}.

Mathematically, Eq.\ (\ref{eq:rho-pi-DAs}) generates a linear relation
between any (linear) functionals depending on the DAs $\varphi_{\rho}^{\|}$
and $\varphi_{\pi}$.
This means that if such a relation holds for the $\rho$-meson
DA, this functional can be replaced by the right hand side (RHS) of Eq.\ (10),
where the $\pi$ DA and $\Phi_S$ enter.

For our further considerations, it is useful to make use
of the convolution
\begin{equation}
  \varphi_\text{M}\otimes f
=
  \int_{0}^{1}\varphi_\text{M}(x)f(x) dx\, .
\label{eq:phi_f}
\end{equation}
Then we can employ the Gegenbauer expansion of the meson DA
$\varphi_\text{M}$ in Eq.\ (\ref{eq:DA-conf-exp}) to get expressions
in terms of the coefficients $a_n$ of meson $M$,
\begin{equation}
 a_{n}^\text{M}=  \varphi_\text{M}\otimes\widetilde{\psi}_n
=
  \int_{0}^{1}\varphi_\text{M}(x)\widetilde{\psi}_{n}(x)dx\, ,
\label{eq:meson-DA-conv}
\end{equation}
where $a_{n}^\text{M}$ are the coefficients of the ``conformal expansion''
over the adjoint harmonics
\begin{equation}
  \widetilde{\psi}_{m}(x)
=
  C_{m}^{3/2}(2x-1)/N_m\,,~ \widetilde{\psi}_{m}\otimes \psi_{n}=\delta_{n m},
\label{eq:adjoint-harmonics}
\end{equation}
and $\Ds N_m=3(m+1)(m+2)/2(2m+3)$ are normalization constants.

Convoluting expression (\ref{eq:rho-pi-DAs}) with
$\widetilde{\psi}_{m}(x)$, we obtain
\begin{eqnarray}
  a_{n}^{(\rho)}
\!&\! \stackrel{\text{sr}}{=} \!&\!
  \left[
        a_{n}^{(\pi)} + \Delta_{A_{1}\rho'}\otimes \widetilde{\psi}_{n}
        - \frac{2}{f_{\pi}^2} \Phi_S\otimes \widetilde{\psi}_{n}
  \right] \nonumber \\
&& \times e^{C(M^2) m_{\rho}^2}
\label{eq:rho-pi-fin}
\end{eqnarray}
in which the last term in the square brackets dominates over the
second one.
For example, for $n=2$, the second ``resonances'' term contributes
only a few $\%$ compared to the scalar NLC.
The notation $\stackrel{\text{sr}}{=}$  means that one should take
the average of the RHS over $M^2$ within the stability window
in the Borel parameter, i.e.,
$$
  0.55~\text{GeV}^2
=
  M^2_{-} < M^2 \lesssim M^2_{+}=1.1~\text{GeV}^2
$$
in order to obtain a certain numerical value of the left hand side (LHS).

\section{Connection between scalar NLC and pion PDF}
\label{sec:NLC-M3}
In this section we work out relation (\ref{eq:rho-pi-fin}) for the
second and fourth meson moment in connection with the NLC-SR and its
vacuum parameters.
To this end, all scale-dependent quantities are evolved from
$\mu^2=4$~GeV$^2$ to the typical scale of QCD SR
$\mu^2_0 \simeq 1$~GeV$^2$ using the NLO evolution procedure described
in \cite{Stefanis:2020rnd}
(see also \cite{Pimikov:2013usa,Stefanis:2015qha}).

\subsection{\!\!\!$\rho^\|-\pi$ relation for the $a_{2}^\text{M}$ moment}
For $n=2$, Eq.\ (\ref{eq:rho-pi-fin}) looks as the analogue of
Eq.\ (\ref{eq:second-moments}) expressed in terms of the Borel-mass
dependent NLC-SRs.
For this reason, also the elements in the RHS of (\ref{eq:rho-pi-fin})
depend on the parameter $M^2$.
Let us now confront the meaning and estimates of the different elements
of this SR with their counterparts in Eq.\ (\ref{eq:second-moments}).
Evaluating the above expression for $n=0$ and supposing that the
normalization conditions
$a_{0}^{(\pi)}=a_{0}^{(\rho)}=1$
hold \emph{within a common window of stability} with respect to $M^2$,
we find
\begin{equation}
  \left[1 + \Delta_{A_{1}\rho'}\otimes \widetilde{\psi}_{0}
        - \frac{2}{f_{\pi}^2} \Phi_S\otimes \widetilde{\psi}_{0}\right]^{-1}
=
 e^{C(M^2) m_{\rho}^2}\, ,
\label{eq:n=0}
\end{equation}
where
$\widetilde{\psi}_{0}=1$.
This equation can be used to estimate the ``constant'' $C(M^2)=C$
which has a physical sense similar to $c_{1}^{(21)}$ in
Eq.\ (\ref{eq:second-moments}).
However, the sum rule (\ref{eq:n=0}) for $C$ is unstable and can only
provide the domain of variation of $C\in [0.8-0.2]$.
Nevertheless, this interval has some overlap with the estimate for the
subtraction constant $c_{1}^{(21)}\in [0.7-0.9]$ calculated in
\cite{Polyakov:2020cnc}.

It is useful to rewrite relation (\ref{eq:rho-pi-fin}) by expressing
the factor $e^{C(M^2) m_{\rho}^2}$ through the LHS of (\ref{eq:n=0})
to obtain the more homogenous form
\begin{subequations}
\label{eq:an-sumrule}
\begin{eqnarray}
\!\!\!\!\!\!\!\!\!\!\!\!\!\!  a_{n}^{(\rho,sr)}
&\stackrel{\text{sr}}{=}&
  \left[
        a_{n}^{(\pi)} - \left[\frac{2}{f_{\pi}^2}\Phi_S -\Delta_{A_{1}\rho'}\right]_n\
  \right]\mathcal{N}(M^2) \label{eq:rho-pi-fin-prime} \\
\!\!\!\!\!\!\!\!\!\!\!\!\!\!\mathcal{N}(M^2)
& \stackrel{\text{sr}}{=}&
  \left[1-\left[\frac{2}{f_{\pi}^2}\Phi_S -\Delta_{A_{1}\rho'}\right]_0 \right]^{-1}\, ,
 \label{eq:prime_n=0}
\end{eqnarray}
\end{subequations}
where
$
 \left[\Delta_{A_{1}\rho'}\right]_n
\equiv
 \Delta_{A_{1}\rho'}\otimes \widetilde{\psi}_{n}
$,
$
 \left[\Phi_S\right]_n
\equiv
 \Phi_S\otimes \widetilde{\psi}_{n}
$,
and $\mathcal{N}(M^2)$ is a normalization factor.
It is worth noting that relation (\ref{eq:rho-pi-fin-prime}) for $n=2$
possesses a sufficient stability with respect to $M^2$ and finally
yields for
$a_{2}^{(\pi)}(\mu^2_0)=0.187$ \cite{Bakulev:2001pa}
(used as input for the evaluation of the RHS) the result
$a_{2}^{(\rho,sr)}(\mu^2_0)=0.047^{+0.035}_{-0.011}$.
This estimate agrees with the value $a_{2}^{(\rho)}(\mu^2_0)=0.047(58)$
obtained from the original NLC-SR in \cite{Pimikov:2013usa} but is more
accurate.

We are now in the position to relate the moment expressions, obtained
within the two considered nonperturbative approaches to the QCD vacuum,
using the scalar condensate.
Evaluating Eq.\ (\ref{eq:an-sumrule}) for $n=2$
and averaging the RHS over $M^2$, we obtain
 \begin{eqnarray}
 a_{2}^{(\rho,sr)}
 \!\!\!\!&=&
  \left[
        a_{2}^{(\pi)}
        - ~\langle\Phi_2 \rangle
 ~\right] \langle\mathcal{N}\rangle, \label{eq:a2SR}\\
 a_{2}^{(\rho)}\!\!\!\!&=&
  \left[
        a_{2} ^{(\pi)} - \frac{7}{6}M_{3}^{(\pi)}
  \right]
   e^{\left(c_{1}^{(21)}m_{\rho}^{2}\right)}\, ,
\nonumber
\end{eqnarray}
where we have shown in the second line Eq.\ (\ref{eq:second-moments})
from \cite{Polyakov:2020cnc} for convenience.
\begin{figure}[h]
\includegraphics[width=0.48\textwidth]{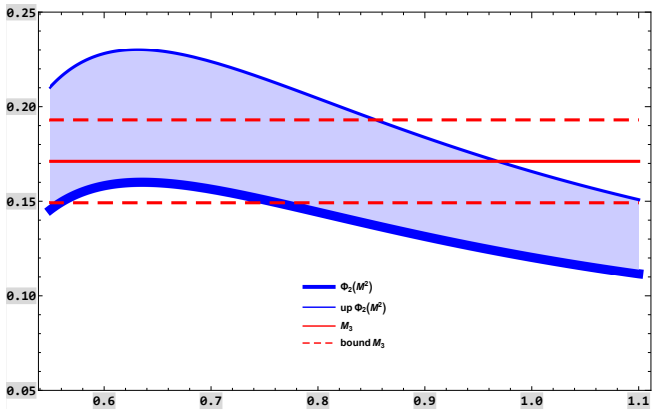}
\caption{
The central solid (red) line represents the mean value of $M_3$ ($y$ axis)
$\frac{7}{6} M_3=0.171(22)$ at $\mu^2_0$
(using the more precise lattice estimate from \cite{Alexandrou:2021mmi}),
while the dashed (red) lines mark its uncertainties
from \cite{Novikov:2020snp}.
The thick (blue) line depending on $M^2$ ($x$ axis) is the RHS of Eq.\ (\ref{eq:final})
before averaging, i.e. $\Phi_2(M^2)$, whereas the upper thinner (blue) line
denotes its upper limit \cite{Cheng:2020vwr} (see Appendix \ref{app:B}).
\label{fig:SR-M3a}}
\end{figure}
Here we employed the notation
\begin{subequations}
\begin{equation}
\Phi_n \equiv \left[\frac{2}{f_{\pi}^2} \Phi_S - \Delta_{A_{1}\rho'}
       \right]_n\frac{\mathcal{N}}{\langle\mathcal{N}\rangle},
\label{eq:Phi_n}
\end{equation}
where $\langle f\rangle$ denotes the average over the stability window of $M^2$:
\begin{eqnarray}
\langle f\mathcal{N}\rangle&=&\int^{M^2_+}_{M^2_-} f(x)\mathcal{N}(x) d\,x
/\left(M^2_+ - M^2_-\right)\, .
\label{eq:average}
\end{eqnarray}
\end{subequations}

In order to get the same outcome from expressions (\ref{eq:a2SR}) and
(\ref{eq:second-moments}), we have to conclude that
\begin{equation}
  \frac{7}{6} M_3
\approx
  \langle \Phi_2\rangle\, .
\label{eq:final}
\end{equation}
Evolving the estimate in (\ref{eq:M3}) to the scale $\mu^2_0$, we
obtain for the LHS of (\ref{eq:final}) the value $0.171(22)$
\cite{Alexandrou:2021mmi}, which is shown as a solid line
in Fig.\ \ref{fig:SR-M3a} together with its uncertainties
denoted by dashed red lines.
The quantity $\Phi_2(M^2)$ together with its upper boundary is displayed
by means of the shaded (blue) strip that has a strong overlap with the
dashed (red) lines. 
The mean value of $\Phi_2(M^2)$ varies in the range
$\langle \Phi_2\rangle = 0.139\div 0.196 $
so that for (\ref{eq:final}) we have
\begin{subequations}
\label{eq:22}
\begin{eqnarray}
\!\!\!\!\text{Eq.\ (\ref{eq:final})}\! &\Rightarrow& \!
\bigg\{
		\begin{array}{l}
		\text{LHS:}\,0.171(22)\!  \\
	\text{RHS:}\,0.139\div 0.196
		\end{array} \label{eq:22a}\\
\text{LHS:}\,0.171(22) &\approx & \text{RHS:}0.167(29),
\label{eq:22b}
\end{eqnarray}
\end{subequations}
where in the RHS of (\ref{eq:22b}) we employed the mean value of the
interval in (\ref{eq:22a}).
Note that the uncertainties in the estimate of $\Phi_2$
are entailed by the uncertainties of the value of the quark condensate
$\langle\sqrt{\alpha_s} \bar{q}q \rangle^2$
and the use of the single parameter $\lambda_q^2$,
cf.\ (\ref{eq:quark-virtuality}), to describe
correlations in the quark NLC, see Appendix \ref{app:A}.

These considerations establish the approximate validity of
Eq.\ (\ref{eq:final}).
This equation represents an intriguing relationship between the
measurable quantity $M_{3}$
(the third moment of the pion PDF) and the
scalar NLC $\Phi_S$ which parameterizes the nontrivial vacuum
of QCD.
Let us recall here that $\Phi_S$ significantly dominates in the RHS
of (\ref{eq:final}), while the LHS of this equation was obtained by
appealing to the general properties of quantum field theory that gave
rise to the nontrivial relation (\ref{eq:second-moments}) derived
in \cite{Polyakov:2020cnc}.
It involves no further assumptions or theoretical modeling.

From this perspective, the condition for the sign of $a_{2}^{(\rho)}$ in
(\ref{eq:ratio}) turns out to be directly related to the QCD vacuum
characteristics in terms of the mean value of the scalar condensate:
\begin{eqnarray}
&&  \text{If}~a_{2}^{(\pi)}(\mu_0^2)
\geqslant
   \langle \Phi_2\rangle(=0.167) \nonumber \\
&&
\text{then}~a_{2}^{(\rho)}(\mu_0^2) \geqslant 0,
\label{eq:23}
\end{eqnarray}
and vice versa.
The same condition can be obtained directly from
Eq.\ (\ref{eq:second-moments}) using the replacement
$\Ds \frac{7}{6}M_3 \rightarrow \langle \Phi_2\rangle$,
as illustrated in Fig.\ \ref{fig:a2-comparison}
(at the scale $\mu^2 =4$~GeV$^2$).

\subsection{Conformal expansion beyond second order}
\label{sec:higher-order}
Let us start with the sum-rule
result in Eq.\ (\ref{eq:rho-pi-fin-prime}) evaluated for $n=4$:
\begin{subequations}
\begin{eqnarray}
\!\!\!\!\!\!  a_{4}^{(\rho,sr)}(\mu_0^2)
\!\!&\!=&\!\!\!\!
  \left[
        a_{4}^{(\pi)}
        - \langle \Phi_4\rangle
  \right]\!\langle\mathcal{N}\rangle\, .
\label{eq:rho-pi-SR-a4}
\end{eqnarray}
From the above equation we can extract the value
$$a_{4}^{(\rho,sr)}= -0.058^{+0.023}_{-0.020}$$
which corresponds to $a_{4}^{(\pi)}= -0.129$
(\ref{eq:rho-pi-SR-a4}) \cite{Bakulev:2001pa,Pimikov:2013usa} in the RHS,
while the value of $\langle \Phi_4\rangle$ turns out to be also negative.
This value of $a_{4}^{(\rho,sr)}$ agrees well (with an even better
accuracy) with
$$a_{4}^{(\rho)}= -0.057(118)$$
obtained in the standard NLC-SR analysis in \cite{Pimikov:2013usa}.
The expressions for the conformal coefficients for any $n$ in both
discussed approaches are given in Appendix \ref{app:B}.

In Table \ref{tab:a24-rho} we present pairs $(a^{\rho}_2,a^{\rho}_4)$
computed by using as input some favored pion DAs:
(i) BMS model \cite{Bakulev:2001pa}
and (ii) a new DA determined in \cite{Mikhailov:2021znq} (denoted there by the symbol
\GrayW{\ding{115}}).
This DA belongs to the BMS family and complies with the latest lattice result
\cite{Bali:2019dqc} at N$^3$LO.
\begin{table}[ht]
\caption{Estimates of
$(a^{\rho}_2,a^{\rho}_4)$ at $\mu_0^2\simeq 1$~GeV$^2$ based on the
results for the \textit{mean values} of $(a^{\pi}_2,a^{\pi}_4)$ within the
NLC-SR \cite{Pimikov:2013usa,Mikhailov:2021znq} and independently on
Eqs.\ (\ref{eq:a2SR}), (\ref{eq:rho-pi-SR-a4}).}
\begin{ruledtabular}
\begin{tabular}{rlll}
meson(\textbf{M}) & source &$\bm{a^\textbf{M}_2}$ &$\bm{a^\textbf{M}_4}$  \\
       \hline
         & NLC-SR\text{\cite{Pimikov:2013usa}}  & $0.047(58)$& $-0.057(0.118)$  \\
$\bm{\rho_\|}$& here (\ref{eq:a2SR}), (\ref{eq:rho-pi-SR-a4})& $0.047^{+0.035}_{-0.011}$&$-0.058^{+0.023}_{-0.020}$  \\
              & here, based on
              & $\bm{\circledast}$ $0.019^{+0.025}_{-0.009}$ & $-0.027$  \\
              & &as input                                    & for $\rho_\|$ DA:  \\
$\bm{\pi}$    & NLC-SR\text{\cite{Bakulev:2001pa,Pimikov:2013usa}} & 0.187 &$-0.129$  \\
              & model $\bm{\circledast}$
         \text{\cite{Mikhailov:2021znq}}      & 0.159 &$-0.098$ \\
\end{tabular}
\end{ruledtabular}
\label{tab:a24-rho}
\end{table}

The dispersive approach elaborated in \cite{Polyakov:1998ze} gives
a linear relation analogous to (\ref{eq:second-moments})
also for the higher conformal coefficients.
Based on the results obtained in \cite{Polyakov:1998ze}
(see, also, the discussion in Appendix \ref{app:B})
one derives the relation
\begin{eqnarray}
\!\!\!\!  a_{4}^{(\rho)}
&\!\! = \!\! &
  \left[
        a_{4} ^{(\pi)} - \frac{11}{9}M_{5}^{(\pi)}
 - B_{43} \right]
  \exp{\left(c_{4}m_{\rho}^{2}\right)} \, ,
\label{eq:fourth-moments}
\end{eqnarray}
\end{subequations}
where the coefficient $B_{43}(0)$ was determined to be
$B_{43}(0)\approx-0.12$,
while the low-energy subtraction constant $c_4$ can only be
poorly estimated from the instanton model \cite{Polyakov:1998ze}
to have the value $c_{4}=1$~GeV$^{-2}$.
Comparing (\ref{eq:fourth-moments}) with (\ref{eq:rho-pi-SR-a4}),
and assuming the approximate equality
\begin{subequations}
\begin{equation}
 \exp{\left(c_{4} m_{\rho}^{2}\right)}
\simeq
 \langle \mathcal{N}\rangle\, ,
\label{eq:exponent}
\end{equation}
we claim the validity of the relation
 \begin{eqnarray}
  \frac{11}{9}M_{5}^{(\pi)}+B_{43}(0)
&= & \langle \Phi_4\rangle
\label{eq:finalM_5}
\end{eqnarray}
 \end{subequations}
The numerical evaluation of the RHS of Eq.\ (\ref{eq:finalM_5})
amounts to the mean value of
$\langle \Phi_4\rangle= - 0.069(50)$
within the fiducial window in $M^2$
(at the normalization scale $\mu^2_0\simeq 1$~GeV$^2$).
In the LHS of this equation we can use the estimate for
$M_{5}^{(\pi)} \equiv\langle x^4 \rangle =0.027(2)$
from the Jefferson Lab Angular Momentum (JAM) PDF in \cite{Alexandrou:2021mmi}
(at $\mu^2=4$ GeV$^2$) and the estimate for $B_{43}(0)$
so that evolving the LHS to $\mu_0^2$, we obtain for
Eq.\ (\ref{eq:finalM_5}) the approximate numerical equality (within errors)
\begin{equation}
\label{eq:26}
\text{LHS:}0.049(4) -0.12\!=\!-0.071(4)\! \approx\! \text{RHS:} -0.069(50)
\end{equation}
to complete the comparison.
The reasonable numerical agreement of the estimates in (\ref{eq:22})
and (\ref{eq:26}) gives support to the claim that the nonlocal scalar
condensate $\Phi_\text{S}$ and the pion non-singlet PDF
$\left[q_\pi(x) - \bar{q}_\pi(x)\right]$
are indeed linearly related to each other.
This unexpected result provides the possibility to unravel the basic
characteristics of the nonperturbative QCD vacuum using measurements in DIS.

Let us now consider the estimate $c_4 = 1$ GeV$^{-2}$
obtained from the instanton model \cite{Polyakov:1998ze}.
For this value, the proposed relation
$\exp{\left(c_{4} m_{\rho}^{2}\right)}\simeq \langle \mathcal{N}\rangle$
in Eq.\ (\ref{eq:exponent}) cannot be realized.
The RHS of Eq.\ (\ref{eq:finalM_5}) becomes instead the weighted sum of
$\langle \Phi_4\rangle$ and $a_{4}^{(\pi)}$ so that
\begin{eqnarray}
\!\!\!\!\!\!\!\!\! \frac{11}{9}M_{5}^{(\pi)}+B_{43}(0)
&\!=\! &\! \langle \Phi_4\rangle \frac{\langle \mathcal{N}\rangle\,}{e^{m^2_{\rho}c_4}} +
a_{4}^{(\pi)}\left(1-\frac{ \langle \mathcal{N}\rangle\,}{e^{m^2_{\rho}c_4}}\right)\, .
\label{eq:finalM_5general}
\end{eqnarray}
Substituting
$a_{4}^{(\pi)}\approx -0.129$ (see Table \ref{tab:a24-rho}) and
$\langle \mathcal{N}\rangle /e^{m^2_{\rho}c_4} \approx 0.54$
into the RHS of this equation, we obtain
\begin{equation}
\label{eq:29}
0.049(4) -0.12\!=\!-0.071(4)\! \sim\!\! \text{RHS:} -0.096(\pm 0.027)\, ,
\end{equation}
where we have neglected the uncertainty of $a_{4}^{(\pi)}$ and the correlation
between the estimates of $B_{43}$ and $c_4$.
From the above considerations, we conclude that Eq.\ (\ref{eq:29}) is
approximately fulfilled even for this case.

\section{Implications for the $\rho^\|$-meson and pion DA}
\label{sec:implications-pion-DA}
Let us now discuss the implications of the cross-link relations derived
in the previous sections for the pion and $\rho^\|$-meson DAs.

Polyakov and Son claimed in \cite{Polyakov:2020cnc} that the different
signs of $a_2^{(\rho)}<0$ and $a_2^{(\pi)}>0$ could be regarded as an
indication that the corresponding DAs may differ significantly.
The origin of this difference could be ascribed to their distinctive
response to the nonperturbative structure of the QCD vacuum.
To make this difference more explicit, we display below their
nonperturbative content in terms of NLC's:
\begin{eqnarray}
&& \Phi_{\rho(\pi)}(x,M^2)
=
  \mp \Phi_\text{S}(x,M^2)
  + \Phi_{\bar q Aq}(x,M^2)
\nonumber \\
&& \quad \quad + \Phi_\text{V}(x,M^2)\! + \Phi_\text{G}(x,M^2) \, .
\label{eq:SRrhoLterms}
\end{eqnarray}
These contributions represent the
(i) $\Phi_{\text{S}}$ (scalar four-quark condensate),
(ii) $\Phi_{\bar q Aq}$ (quark-gluon-antiquark condensate),
(iii) $\Phi_\text{V}$ (vector quark condensate), and
(iv) $\Phi_\text{G}$ (gluon condensate),
with explicit expressions given in Appendix A of \cite{Mikhailov:2010ud}.
One notices that the four-quark condensate enters the expression for the
$\rho^\|$-meson DA with the opposite sign relative to the $\pi$ DA.
As a consequence, it tends to reduce the relative weight of this
condensate with the result that the DA moments (or conformal coefficients)
become smaller.
This may indeed entail different shapes for these DAs.

\begin{figure}[t]
\includegraphics[width=0.48\textwidth]{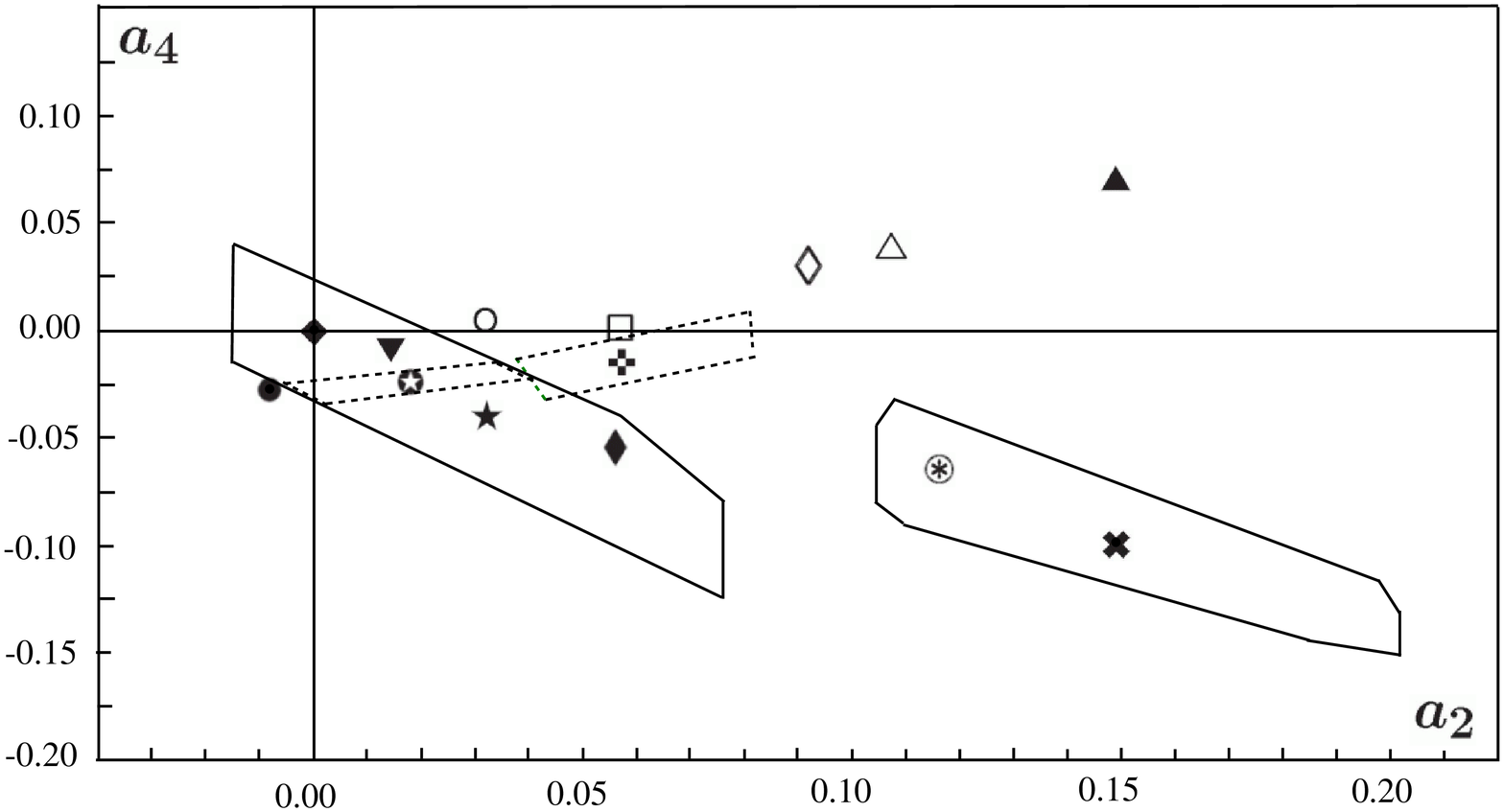}
\caption{Snapshot of various $\pi$ and $\rho^\|$ DAs from different
approaches at the scale $\mu=4$~GeV$^2$ in the $(a_2,a_4)$ plane.
The large ``rectangle'' intersecting with the $a_2=0$ line denotes
the region of $\rho^\|$ DAs determined with NLC-SR.
It encloses the shaded platykurtic regime, where \ding{74} is the DA
determined in \cite{Stefanis:2015qha}.
The symbol \ding{72} refers to the bimodal DAs from NLC-SR
in \cite{Pimikov:2013usa} and
$\blacklozenge$ \cite{Bakulev:1998pf}, respectively.
The symbol \ding{116} represents the $\rho^\|$ DA from the lightfront
model in \cite{Choi:2007yu} and \ding{117} marks the asymptotic DA.
Outside this ``rectangle'' one has the following $\rho^\|$ DAs:
$\square$ \cite{Ahmady:2012dy} and $\bm{\lozenge}$ \cite{Gao:2014bca}.
The other large ``rectangle'' further to the right contains the domain
of the BMS pion DAs \cite{Bakulev:2001pa} obtained with NLC-SR, with
\ding{54} denoting the bimodal BMS pion DA.
The pion DA represented by $\bm{\circledast}$ was determined in
\cite{Mikhailov:2021znq} as the crossing point of the long middle
line of this ``rectangle'' with the lattice result
$a_{2}^{(\pi)}=0.116^{+19}_{-20}$ at N$^3$LO with three-loop matching
to the \MSbar scheme \cite{Bali:2019dqc}.
The dashed ``rectangle'', crossing the $a_{4}^{(\pi)}=0$ line, shows
the platykurtic range of pion DAs with \ding{60} marking the DA
determined in \cite{Stefanis:2014nla}.
The symbols $\bigtriangleup$ and \ding{115} reproduce, respectively,
the DAs obtained within holographic AdS/QCD \cite{Brodsky:2007hb} and a
DSE-based approach \cite{Chang:2013pq}, while $\circ$ shows the pion DA
from the instanton model in \cite{Petrov:1998kg} after NLO evolution to
the scale 4~GeV$^2$.
Similarly, \ding{108} represents the instanton-based $\rho^\|$ DA from
\cite{Dorokhov:2006xw}.
\label{fig:a2-a4-space}}
\end{figure}

For instance, the $\rho^\|$ DA could have a unimodal shape close to the
asymptotic form describing a more or less equal distribution of longitudinal
momentum between the two valence quarks, while the nonperturbative effects
in the pion could lead to a much broader unimodal or bimodal distribution which
favors unequally distributed momentum shares away from $x=1/2$,
see, e.g., \cite{Chernyak:1983ej,Bakulev:2001pa}.

A negative sign prediction for $a_2^{(\rho)}$ is not completely new.
In fact, in \cite{Stefanis:2015qha} it was found that in the case of the
two $\rho^\|$ DAs determined with NLC-SR, notably the platykurtic
DA \cite{Stefanis:2015qha} and the DA from \cite{Pimikov:2013usa}, the range of
$a_2^{(\rho)}$ \emph{can} indeed be negative, see Fig.\ \ref{fig:a2-a4-space}.
However, this possibility was not considered any further.
In contrast, the negative sign of $a_2^{(\rho)}$ in \cite{Polyakov:2020cnc}
appears as a strict finding of rather general principles of QFT and
the instanton model, giving rise to the condition
$a_{2}^{(\pi)}(\mu^2=4~\text{GeV}^2)= 0.078< 7/6\, M^{(\pi)}_3$,
and, thus, deserves particular attention.

Let us discuss these issues in terms of Fig.\ \ref{fig:a2-a4-space},
which shows the locations of various pion and $\rho^\|$ DAs
in the $(a_{2},a_{4})$ plane at the scale $\mu^2=4$~GeV$^2$, and apply
Eq.\ (\ref{eq:second-moments})
in conjunction with Fig.\ \ref{fig:a2-comparison}.
Then, only two classes of $\rho^\|$ DAs are possible depending on the
sign of $a_{4}^{(\rho)}$.
If it is positive, then the corresponding DA will belong to the area
above the $a_{4}^{(\rho)}=0$ line and will have a profile close to the
asymptotic DA.
If $a_4$ is negative, the DA will be located below the
$a_{4}^{(\rho)}=0$ line and will have an almost asymptotic profile
combined with a mild endpoint suppression.
The broadest variant of these $\rho^\|$ DAs is a platykurtic, i.e., a
unimodal distribution with suppressed endpoints \cite{Stefanis:2015qha},
belonging to the embedded dashed ``rectangle'' in
Fig.\ \ref{fig:a2-a4-space}.

In all considered cases, these DAs will differ substantially from the
endpoint-suppressed, but bimodal pion DAs within the BMS ``rectangle''
\cite{Bakulev:2001pa} shown in this figure in terms of solid lines.
The difference between a quasi-asymptotic $\rho^\|$ DA and the
endpoint-enhanced bimodal Chernyak-Zhitnitsky pion DA \cite{Chernyak:1983ej},
obtained with local condensates with $a_2^\text{CZ}(2~\text{GeV})=0.42$
\cite{Stefanis:2015qha} is even stronger and is located outside the
displayed range of $a_{2}^{(\pi)}$.
Similar considerations apply also to broad unimodal pion DAs as those
obtained from the DSE based approach,
see \cite{Roberts:2021nhw} for a recent review.
Indeed, from Fig.\ \ref{fig:a2-a4-space} one sees that the corresponding
pion DA, given by \ding{115}, is a broad unimodal distribution differing
strongly from the asymptotic one.
A similar observation applies also to the pion DA derived in holographic
QCD, shown by $\bigtriangleup$.
Hence, this scenario supports the assumption that the $\pi$
and the $\rho^\|$ interact differently with the nonperturbative QCD
vacuum giving rise to very different DAs.

The other possible scenario would allow both meson DAs to have similar
shapes.
In fact, the $\rho^\|$-meson DA can be obtained from the pion bilocal
correlator by the replacement
$\gamma_\mu\gamma_5\rightarrow \gamma_\mu$
\cite{Chernyak:1983ej} so that one may naively think that this does not
cause significant changes.
The simplest possibility would be that their shapes are both close to
$\varphi_{\pi}^\text{asy}$, see Fig.\ 1 in \cite{Stefanis:2015qha}.
But it has been shown by various authors that in this case, the
pion-photon transition form factor will underestimate the data
considerably, see, for instance, \cite{Stefanis:2020rnd} for a
recent state-of-the-art analysis based on lightcone sum rules and
further references.
This implies that the pion DA must be broader than the asymptotic form
at experimentally accessible momenta.

The only remaining possibility for these mesons to have similar
profiles with $a_2^{(\rho)}<0$ and $a_2^{(\pi)}>0$ in combination with
pion-photon TFF predictions that agree with most experimental data, is
that both have $a_4^{(\rho)}<0$ and $a_4^{(\pi)}<0$.
In this case, their corresponding profiles will have to be platykurtic.
As has been shown in \cite{Stefanis:2014nla,Stefanis:2015qha}, and also
more recently in \cite{Mikhailov:2021znq}, the platykurtic pion DA
reproduces the trend of all data supporting asymptotic scaling at high
$Q^2$ from $Q^2\geq 1$~GeV$^2$ up to momenta $\sim 40$~GeV$^2$.
The most appropriate numerical values of the conformal coefficients, can be
extracted by imposing additional constraints from lattice simulations.
It is worth noting that the latest lattice calculation of $a_2^{(\rho)}$
in \cite{Braun:2016wnx} gives the rather large positive value
$0.132(\pm 0.027)$ that is incompatible with the restriction $a_2^{(\rho)}<0$.
On the other hand, employing the most recent lattice constraint on
$a_2^{(\pi)}$ with NLO or NNLO accuracy from \cite{Bali:2019dqc}, one
finds from Fig.\ \ref{fig:a2-comparison} that
$a_{2}^{(\rho)}\in [-0.01\div -0.09]$.
These lattice constraints can provide in combination with the pion-photon
TFF data best-choice parameters
$a_2^{(\pi)}$ and $a_4^{(\pi)}$,
from their overlapping region as shown in \cite{Stefanis:2020rnd}.
However, the new $a_2^{(\pi)}=0.116_{-0.020}^{+0.019}$ lattice estimate
of \cite{Bali:2019dqc} at N$^3$LO with three-loop matching to the \MSbar
scheme favors a pion DA with a moderate bimodal profile, like
$\bm{\circledast}$ \cite{Mikhailov:2021znq}, giving support to the first
considered scenario.
Further constraints are needed along the upper diagonal in
Fig.\ \ref{fig:a2-a4-space} in order to resolve the fine details of
the pion DA more reliably.

\section{Conclusions}
\label{sec:concl}
In this work we investigated cross-link relations between the Gegenbauer
coefficients of the $\rho^\|$-meson and pion DAs.
We showed that the linear relation between $a_{2}^{(\rho)}$ and
$a_{2}^{(\pi)}$ obtained recently in \cite{Polyakov:2020cnc}
on the basis of rather general assumptions in combination with estimates
from the instanton vacuum can also be obtained using QCD sum rules with
nonlocal condensates.
In fact, we were able to derive an intriguing relation between the third
Melin moment of the pion PDF measured in DIS and the
scalar condensate (see Eq. (\ref{eq:final})).
We also extended this cross-link relation to the next order Gegenbauer coefficients
establishing further the connection between these two nonperturbative descriptions
of the QCD vacuum (see Eq.\ (\ref{eq:finalM_5})).
These findings may contribute to a better understanding of the intrinsic structure
of the QCD vacuum by measurements of the pion PDF.
The COMPASS++/AMBER experiment at CERN may provide high-precision data for
the pion structure function to extract such information.

Adopting a broader perspective, we discussed the general implications for the
pion and $\rho^\|$-meson DAs entailed by the strict application of
Eq.\ (\ref{eq:ratio}) by formulating two different generic scenarios.
With respect to the space $(a_2,a_4)$, we found that the imposition of
$a_{2}^{(\rho)}<0$ not only reduces the available range of the $\rho^\|$-meson
coefficients, it also contributes important constraints for the proper
selection of the pion DA.
We concluded that either these mesons may have (for whatever reason) very
different DAs or, if these are assumed to be similar (for whatever reason),
then they can only have a platykurtic, i.e., unimodal profile with
suppressed tails
\cite{Stefanis:2014nla,Stefanis:2014yha,Stefanis:2015qha,Stefanis:2020rnd}.
The most recent lattice result from \cite{Bali:2019dqc} for $a_{2}^{(\pi)}$
seems to support at N$^3$LO rather the first scenario while the analogous
estimates at NLO and NNLO conform also with the second option.
The possibility that \emph{both} DAs are close to the asymptotic form seems
to be excluded because the pion-photon TFF calculated with
$\varphi_{\pi}^\text{asy}$ underestimates the available data considerably
demanding a broader distribution for the pion DA.

\acknowledgments
We would like to thank Maxim Polyakov for inspiring discussions and
valuable comments on the manuscript.
We have also profited from conversations with Hyeon-Dong Son and 
Andrei L.\ Kataev.
S.~V.~M. acknowledges support from the Heisenberg--Landau Program 2021.

\begin{appendix}
\appendix
\section{Nonlocal scalar condensate}
\label{app:A}
Application of the factorization ansatz on the four-quark condensate,
leads to the product of a pair of scalar condensates $M_S$,
where
$
 M_S(z^2)
=
 \int_0^\infty \exp\left(-z^2/4  \alpha \right) f_{S}(\alpha) d\alpha
$.
The correlation functions $f_{S}(\alpha)$ for each of these scalar
condensates determine the distribution in virtuality $\alpha$ of the
quarks in the QCD vacuum \cite{Bakulev:2002hk,Mikhailov:2010ud}.
For example, for
$f_{S}(\alpha;\Lambda,\sigma) \sim \alpha^{n-1} e^{-\Lambda^2/\alpha-\alpha\,\sigma^2}$
one obtains the expected exponential asymptotic behavior for $M_S(z^2)$ at large $z^2$:
$M_S(z^2) \sim \exp{\left(-\Lambda z \right)}$.

In the NLC approach the factorization ansatz may lead to an
overestimation of the four-quark condensate contribution
$\Phi_S\left(x;M^2\right)$
because it evidently neglects the correlation between these pairs.
The relevant expression within this approximation is given by
\begin{widetext}
 \begin{eqnarray}\label{eq:Delta.Phi.S}
  \Phi_S\left(x,M^2\right)
&=&
  \frac{18A_S}{M^4}
  \mathop{\int\!\!\!\!\int}^{~~\infty}_{0\,0~}\!\!
  d\alpha_1\, d\alpha_2\,
  f_S(\alpha_1)\,
  f_S(\alpha_2)\,
\\\nonumber
&\times&
  \frac{x\,\theta \left(\Delta_1-\bar{x}\right)}
  {\Delta _1^2 \Delta _2 \bar{\Delta }_1^2}
  \left[\bar{x}\Delta_2\bar{\Delta}_1
  +\ln\left(\frac{x\Delta_1\bar{\Delta}_2}
  {x\Delta _1-(\Delta_1-\bar{x})\Delta_2}
  \right)
        \Delta_1(\Delta_1-\bar{x})\bar{\Delta}_2
  \right]
+ (x \to \bar{x}) \ ,
\end{eqnarray}
\end{widetext}
referring for further details to Appendix A in \cite{Mikhailov:2010ud}.
The following notations are used:
$
 \Ds A_S
=
 \left(8\pi/81\right)\langle\sqrt{\alpha_s}\bar{q}(0)q(0)\rangle^2
$
and
$
 \langle\sqrt{\alpha_s}\bar{q} q\rangle^2
=
 \left(~1.84^{+0.84}_{-0.24}~\right)\times 10^{-4}$\,GeV$^6$,
$\Delta_i \equiv \alpha_i/M^2$, where
$\bar\Delta_i \equiv 1-\Delta_i$, and $\bar x\equiv 1-x$.
In this work, we used for the NLC estimates the simplest delta-function ansatz
$
 f_{S}(\alpha)
=
 \delta(\alpha - \lambda_q^2/2)
$,
proposed in \cite{Mikhailov:1988nz,Mikhailov:1991pt,Bakulev:1991ps},
it is sufficient for the accuracy of the moment QCD SRs.
This model leads to a Gaussian decay of the scalar quark condensate
$M_S(z^2)$ with $\Delta \equiv \lambda_q^2/(2M^2)$:
\begin{widetext}
\begin{eqnarray}
 \label{qs}
 \Phi_S \left(x;M^2\right)
  &=&\frac{A_S}{M^4}
      \frac{18}{\bar\Delta\Delta^2}
       \Bigl\{
        \theta\left(\bar x>\Delta>x\right)
         \bar x\left[x+(\Delta-x)\ln\left(\bar x\right)\right]
         +(\bar{x} \rightarrow x)
          \nonumber \\
&&\qquad\qquad  +
\theta(1>\Delta)\theta\left(\Delta>x>\bar\Delta\right)
         \left[\bar\Delta
               +\left(\Delta-2\bar xx\right)\ln(\Delta)\right]
         \Bigr\}\ .
\label{eq:phi_s}
\end{eqnarray}
\end{widetext}
A finite value of $\Delta$, related to
the ``decay rate'' of the correlation length of the NLC
\cite{Mikhailov:2010ud}, shifts the weight of the nonperturbative
contributions away from the endpoints.
On the other hand, for $\lambda_{q}^2=0$, all nonperturbative
contributions are concentrated just at the endpoints.

\section{Relations between the conformal coefficients
$a_{n}^{(\rho)}$ and $a_{n}^{(\pi)}$}
\label{app:B}
We provide here the expressions relating $a_{n}^{(\rho)}$ to $a_{n}^{(\pi)}$. \\
(i) Within the framework of QCD sum rules (label sr) with nonlocal condensates
\cite{Bakulev:1998pf,Bakulev:2001pa}:
\begin{eqnarray}
\!\!\!\!\!\!\!\!\!\!\!\!\!\!  a_{n}^{(\rho,sr)}
&=&
  \left[
        a_{n}^{(\pi)} -\langle\Phi_n\rangle
  \right]\langle\mathcal{N}\rangle
\label{eq:rho-pi-fin-prime-2}\,.
\end{eqnarray}
(ii) A similar relation can be obtained within the dispersive approach.
Indeed, in \cite{Polyakov:1998ze,Polyakov:2020cnc} the coefficient
$a_{n}^{(\pi)}$ of the conformal
expansion was expressed as a sum
\begin{equation}
\label{eq:9}
a_{n}^{(\pi)} = \sum_{l=1(\text{odd})}^{n+1} B_{n l}(0)\, .
\end{equation}
Extracting in the sum in the RHS  the extreme terms for $l=1~\text{and}~l=n+1$\, ,
notably,
\begin{equation}
  B_{n 1}(0)=a_{n}^{(\rho)} \exp{\left(- c_n m^2_{\rho}\right)}\, ,
  B_{n n+1}(0) = p(n)M^{(\pi)}_{n+1}\, ,
\nonumber
\end{equation}
one arrives at the expression
\begin{eqnarray}
\!\!\!\!\!\!\!\!\!\!\!\!\!\!  a_{n}^{(\rho)}
&=&
  \left[
        a_{n}^{(\pi)} - p(n)M^{(\pi)}_{n+1} -\sum_{l=3}^{n} B_{nl}(0)
  \right]
         e^{c_n m^2_\rho}\label{eq:rho-pi-fin-prime-0}\, ,\\
&&  p(n)=\frac{3(n+1)}{N_n}=\frac{2(2n+3)}{3(n+2)} \nonumber
\end{eqnarray}
that can be compared with Eq.\ (\ref{eq:rho-pi-fin-prime-2}).
\end{appendix}


\end{document}